
\input phyzzx.tex

\def\np{Nucl. Phys.}
\def\pl{Phys. Lett.}

\def\cmp{Comm. Math. Phys.}

\def\mpl{Mod. Phys. Lett.}

\def\phyrep{Phys. Rep.}

\tolerance=500000
\overfullrule=0pt
\Pubnum={US-FT-11/94}
\pubnum={US-FT-11/94}
\date={July, 1994}
\pubtype={}
\titlepage

\title{Superconformal current algebras and topological
field theories}
\author{J. M. Isidro
\foot{On leave from : Departamento de F\'\i sica de Part\'\i culas,
Universidad de Santiago,
E-15706 Santiago de Compostela, Spain.} }
\address{Department of Physics, Queen Mary  and Westfield College,\break
Mile End Road, London E1 4NS, United Kingdom.}
\author{A.V. Ramallo }
\address{Departamento de F\'\i sica de
Part\'\i culas, \break Universidad de Santiago, \break
E-15706 Santiago de Compostela, Spain.}

\abstract{ Topological conformal field theories based on superconformal
current algebras are constructed. The models thus obtained are the
supersymmetric version of the $G/G$ coset theories. Their topological
conformal algebra is generated by operators of dimensions $1$, $2$ and
$3$ and  can be regarded as an extension of the twisted $N=2$
superconformal algebra. These models possess an extended supersymmetry
whose generators are exact in  the topological BRST
cohomology. }

\endpage
\pagenumber=1
\sequentialequations
\hyphenation {o-pe-ra-tor}

Topological field theories
\REF\wittop{E. Witten \journal\cmp&117(88)353.}
\REF\bbrt{For a review see D. Birmingham, M. Blau, M.Rakowski
and G. Thompson \journal\phyrep&209(91)129.}
[\wittop,\bbrt]
are quantum field theories that do not
possess any local degree of freedom. They are endowed with a topological
symmetry that allows to eliminate all local excitations, in such a way
that only the collective modes of the basic fields  remain in
its spectrum after the topological symmetry is fixed. Some models can be
converted into  topological theories by adding a ghost sector and a BRST
symmetry that can be used to define a cohomology. In so doing one expects
to obtain some non-trivial information about the non-perturbative sector
of the initial theory. We have recently
\REF\tca{J. M. Isidro and A. V. Ramallo\journal\pl&B316(93)488.}
\REF\gln{J. M. Isidro and A. V.
Ramallo\journal\np&B414(94)715.}[\tca,\gln]
applied this procedure to generate
two-dimensional topological conformal field theories based on non-abelian
current algebras. In a model with a current algebra symmetry, the local
degrees of freedom are created by acting with the currents on the vacuum.
The BRST symmetry studied in [\tca,\gln] is such that the total Kac-Moody
current is BRST-exact (\ie, cohomologically trivial). Moreover, we have
shown that the topological conformal algebra contains operators of
dimensions $1$, $2$ and $3$ and can be considered as an extension of the
twisted
\REF\EY{T. Eguchi and S.-K. Yang \journal\mpl&A4(90)1653;
T. Eguchi, S. Hosono and S.-K. Yang \journal\cmp&140(91)159.}[\EY]
$N=2$ superconformal algebra
\REF\LVW{W. Lerche, C. Vafa and N.P. Warner
\journal\np&B324(89)427.}[\LVW].

In this paper we propose to construct topological theories based on
superconformal current algebras
\REF\DVKPR{P. Di Vecchia, V.G. Knizhnik, J.L. Petersen and P.
Rossi \journal\np&B253(85)701.}
\REF\KT{V.G. Kac and I.T. Todorov\journal\cmp&102(85)337.}[\DVKPR,\KT].
Let us consider a finite-dimensional, semisimple Lie
algebra $g$, generated by the hermitian matrices $T^a$
$(a=1,\ldots, {\hbox{\rm dim}}\,g$). The  $T^a$ are chosen such that
Tr($T^aT^b$ )=$\delta_{ab}$ and to satisfy the commutation relations
$$
[T^a,T^b]=if^{abc}T^c.
\eqn\uno
$$
A superconformal current algebra is generated by a set of holomorphic
bosonic currents ${\cal J}_a(z)$ and their fermionic partners $\Phi_a(z)$,
satisfying the following operator product expansions (OPE's):

$$
\eqalign{
{\cal J}_a(z_1){\cal J}_b(z_2)=&{x\delta_{ab}\over (z_1-z_2)^2}
+if^{abc}\,\,\,{{\cal J}_c(z_2)\over z_1-z_2}\cr
{\cal J}_a(z_1)\Phi_b(z_2)=&if^{abc}\,\,\,{\Phi_c(z_2)\over z_1-z_2}\cr
\Phi_a(z_1)\Phi_b(z_2)=&{x\delta_{ab}\over z_1-z_2}\,\, ,\cr}
\eqn\dos
$$
where $x$ is the level of the current algebra. ${\cal J}_a$ and $\Phi_a$
are primary fields with respect to the energy-momentum tensor $T$ with
conformal dimensions $1$ and ${1\over 2}$ respectively. This
model is supersymmetric, which implies that  a
dimension-${3\over 2}$ fermionic operator $T_F$ exists with the OPE
$$
T_F(z_1)T_F(z_2)={c\over 6(z_1-z_2)^3}+{1\over 2}\,\,\,
{T(z_2)\over z_1-z_2}\,\, ,
\eqn\tres
$$
where $c$ is the central charge of the Virasoro algebra. The currents
${\cal J}_a$ and $\Phi_a$ form a doublet under the supersymmetry algebra
generated by $T_F$, so  we must have
$$
\eqalign{
T_F(z_1)\Phi_a(z_2)=&{1\over 2}\,\,\,{{\cal J}_a(z_2)\over z_1-z_2}\cr
T_F(z_1){\cal J}_a(z_2)=&{1\over 2}\,\,\,{\Phi_a(z_2)\over (z_1-z_2)^2}+
{1\over 2}\,\,\,{\partial \Phi_a(z_2)\over z_1-z_2}\,\, .\cr}
\eqn\cuatro
$$
It is possible to realize this superconformal algebra in terms of a
system of uncoupled bosonic currents $j_a$ and dim$\,g$ free Majorana
fermions $\psi_a$  transforming in the adjoint representation of $g$,
the corresponding OPE's being
$$
\eqalign{
j_a(z_1)j_b(z_2)=&{k\delta_{ab}\over
(z_1-z_2)^2}+if^{abc}\,\,{j_c(z_2)\over z_1-z_2}\cr
\psi_a(z_1)\psi_b(z_2)=&-{\delta_{ab}\over z_1-z_2}\,\,.\cr}
\eqn\cinco
$$
The energy-momentum tensor $T$ is obtained by combining the
Sugawara form of the bosonic sector with the canonical energy-momentum
tensor of the Majorana fermions\footnote*{In the following,
although we will not denote it
explicitly, all products of fields will be
understood as normal-ordered.},
$$
T^{(j,\psi)}={1\over 2(k+h)}\,\,j_aj_a+{1\over 2}\,\,\,
\psi_a\partial\psi_a\,\, ,
\eqn\seis
$$
where $h$ is the dual Coxeter number of $g$. For simply-laced algebras
$h={\hbox{\rm dim}}\,g/ {\hbox{\rm rank}}\,g\,-1$ (\ie, for
instance, $h=N$ for $g=sl(N)$). $T^{(j,\psi)}$ closes a
Virasoro algebra with central charge
$$
c^{(j,\psi)}={k\,{\hbox{\rm dim}}\,g\over k+h}+
{1\over 2}\,\,\, {\hbox{\rm dim}}\,g\,\, .
\eqn\siete
$$
Now it's easy to check that the supersymmetry generator $T_F$ satisfying
\tres\ is
$$
T^{(j,\psi)}_F={i\over 2(k+h)^{1\over 2}}\,\,\,j_a\psi_a
-{1\over 12 (k+h)^{1\over 2}}\,\,\,f^{abc}\psi_a\psi_b\psi_c\,\, ,
\eqn\ocho
$$
and that the currents
$$
\eqalign{
{\cal J}_a^{(j,\psi)}=&j_a+{i\over 2}f^{abc}\,\,\,\psi_b\psi_c\cr
\Phi_a^{(j,\psi)}=&i(k+h)^{1\over 2}\psi_a\cr}
\eqn\nueve
$$
close the superconformal current algebra \dos\ with level
$$
x^{(j,\psi)}=k+h\,\, .
\eqn\diez
$$
Moreover it can be readily verified that ${\cal J}_a$ and $\Phi_a$ form a
doublet with respect to $T_F^{(j,\psi)}$, \ie\ ${\cal J}_a$ and $\Phi_a$ as
given by eq. \nueve\ satisfy \cuatro.

The superconformal current algebra \dos\ can also be realized in a
supersymmetric ghost system. Let us introduce a pair of anticommuting
ghosts $\gamma_a$ and $\rho_a$ with dimensions $0$ and $1$ respectively.
The supersymmetry requirement leads us to introduce commuting ghosts
$\eta_a$ and $\lambda_a$, both of them with dimension ${1\over 2}$.
To the fields $\gamma_a$ and $\eta_a$ ($\rho_a$ and $\lambda_a$) we shall
assign ghost number $+1$($-1$ respectively).
Let us
 choose our conventions in such a way that the ghost fields  obey
the OPE's
$$
\eqalign{
\rho_a(z_1)\gamma_b(z_2)=&-{\delta_{ab}\over z_1-z_2}\cr
\eta_a(z_1)\lambda_b(z_2)=&-{\delta_{ab}\over z_1-z_2}\,\, .\cr}
\eqn\once
$$
Since the canonical energy-momentum tensor of this system is
$$
T^{(gh)}=\rho_a\partial\gamma_a+{1\over 2}\,\,\,
\partial\eta_a\lambda_a -{1\over 2}\,\,\, \eta_a\partial \lambda_a
\,\, ,\eqn\doce
$$
the $(\gamma ,\rho)$ and $(\eta ,\lambda)$ systems contribute
respectively with $-2\,$dim$\,g$ and $-$dim$\,g$ to the Virasoro central
charge. Therefore their conformal anomaly is
$$
c^{(gh)}=-3\,\,{\hbox{\rm dim}}\,\,\,g\,\,.
\eqn\trece
$$
The supersymmetry in this ghost system is generated by
$$
T_F^{(gh)}={1\over 2}\eta_a\rho_a-{1\over
2}\partial\gamma_a\lambda_a\,\, ,
\eqn\catorce
$$
and the explicit form of the superconformal currents is
$$
\eqalign{
{\cal J}_a^{(gh)}=&if^{abc}\gamma_b\rho_c+if^{abc}\lambda_b\eta_c\cr
\Phi_a^{(gh)}=&if^{abc}\gamma_b\lambda_c\, .\cr}
\eqn\quince
$$
In this case the algebra \dos\ is closed with vanishing level, \ie, one
has
$$
x^{(gh)}=0\,\, .
\eqn\dseis
$$

Let us now combine the matter and ghost realizations we have described
above. Suppose we consider $M$ copies of the matter system, together with
one copy of
the supersymmetric ghost model, and let us denote by  $j_a^l$ and $\psi_a^l$
($l=1,\ldots, M$) the basic fields of the matter sector. The currents
of the combined matter + ghost system are given by
$$
\eqalign{
{\cal J}_a=&\sum_{l=1}^M(j_a^l+{i\over 2}f^{abc}\,\,\,\psi_b^l\psi_c^l)+
if^{abc}\gamma_b\rho_c+if^{abc}\lambda_b\eta_c\cr
\Phi_a=&i\sum_{l=1}^M(k_l+h)^{1\over
2}\psi_a^l+if^{abc}\gamma_b\lambda_c\,\, ,\cr}
\eqn\dsiete
$$
where $k_l$ is the level of the current $j_a^l$. The complete energy-momentum
tensor now takes the form
$$
T=\sum_{l=1}^M{1\over 2(k_l+h)}\,\,j_a^lj_a^l+
{1\over 2}\sum_{l=1}^M\,\psi_a^l\partial\psi_a^l+
\rho_a\partial\gamma_a+{1\over 2}\,\,\,
\partial\eta_a\lambda_a -{1\over 2}\,\,\, \eta_a\partial \lambda_a\,\, ,
\eqn\docho
$$
and the central charge $c$ and level $x$ are given by
$$
\eqalign{
c=&\sum_{l=1}^M{k_l\,{\hbox{\rm dim}}\,g\over k_l+h}+
({M\over 2}-3)\,\,\, {\hbox{\rm dim}}\,g\cr
x=&\sum_{l=1}^Mk_l+M\, h\, .\cr}
\eqn\dnueve
$$
We would like  this combined matter + ghost system to be a topological
theory. As it was shown in [\tca], a way of finding the appropriate content of
the matter sector is to impose  the zero level condition $x=0$. If we had
only one matter species (\ie, for $M=1$), $x=0$ would mean $k_1+h=0$ (see eq.
\dnueve). However the value $k_1=-h$ is not acceptable since in this case the
expression of the energy-momentum tensor $T$ in \docho\ becomes singular. Let
us therefore consider the case $M=2$. With two currents in the
matter sector, the condition $x=0$ implies
$$
k_1+k_2=-2h\, .
\eqn\veinte
$$
An indication that we are now pointing in the right direction is the fact
that the central charge $c$ vanishes when eq. \veinte\ holds
(see \dnueve), which confirms that we have found a topological point of
the matter + ghost system.
In fact, when the two levels $k_1$ and $k_2$ are related
as in \veinte, the $N=1$ supersymmetry is extended to a larger algebra
that we are now going to describe. The generators of this extended
algebra are naturally expressed in terms of the following complex
combinations of the two real Majorana fields $\psi_a^1$ and $\psi_a^2$:
$$
\Psi_a={1\over \sqrt 2}(\psi_a^1+i\psi_a^2)
\,\,\,\,\,\,\,\,\,\,\,\,
\overline\Psi_a={1\over \sqrt 2}(\psi_a^1-i\psi_a^2)\,\, ,
\eqn\vuno
$$
so from \cinco\ one gets
$$
\eqalign{
\Psi_a(z_1)\Psi_b(z_2)=&\overline\Psi_a(z_1)\overline\Psi_b(z_2)=0\cr
\Psi_a(z_1)\overline\Psi_b(z_2)=&-{\delta_{ab}\over z_1-z_2}\,\, .\cr}
\eqn\vdos
$$
Similarly, let us also define the combinations of the bosonic
currents
$$
J_a=j_a^1+j_a^2
\,\,\,\,\,\,\,\,\,\,\,\,
\bar J_a=j_a^1-j_a^2\,\, ,
\eqn\vtres
$$
whose OPE's are
$$
\eqalign{
J_a(z_1)J_b(z_2)=&\bar J_a(z_1)\bar J_b(z_2)=
-{2h\over (z_1-z_2)^2}\,\,\delta_{ab}+
if^{abc}\,\,{J_c(z_2)\over z_1-z_2}\cr
J_a(z_1)\bar J_b(z_2)=&{2(\kappa+h)\over (z_1-z_2)^2}\,\,\delta_{ab}+
if^{abc}\,\,{\bar J_c(z_2)\over z_1-z_2}\,\,.\cr}
\eqn\vcuatro
$$
We have written $\kappa$ instead of $k_1$ (and therefore
$k_2=-2h-\kappa$). As $(k_2+h)^{{1\over 2}}=i(\kappa+h)^{{1\over 2}}$,
in terms of these new variables the superconformal
currents take the form
$$
\eqalign{
{\cal J}_a=&J_a+if^{abc}\,\,\overline \Psi_b\Psi_c+
if^{abc}\,\, \lambda_b\eta_c+if^{abc}\,\, \gamma_b\rho_c\cr
\Phi_a=&i\sqrt{2(\kappa+h)}\,\,\Psi_a+if^{abc}
\gamma_b\lambda_c\,\, .\cr}
\eqn\vcuatrobis
$$

Consider now the following dimension-${3\over 2}$
operators:
$$
\eqalign{
T_F^{+}=&{1\over \sqrt{2(\kappa+h)}}(iJ_a\,\overline\Psi_a-
{1\over
2}f^{abc}\,\,\overline\Psi_a\,\overline\Psi_b\,\Psi_c)+\eta_a\rho_a\cr
T_F^{-}=&{i\over \sqrt{2(\kappa+h)}}\bar
J_a\Psi_a-\partial\gamma_a\lambda_a \,\, ,\cr}
\eqn\vcinco
$$
which, as  can be easily checked,  are primary with respect to $T$ and
satisfy the algebra
$$
\eqalign{
T_F^{+}(z_1)T_F^{+}(z_2)=&0\cr
T_F^{+}(z_1)T_F^{-}(z_2)=&{R_F(z_2)\over (z_1-z_2)^2}+
{T(z_2)+{1\over 2}\partial R_F(z_2)\over z_1-z_2}\cr
T_F^{-}(z_1)T_F^{-}(z_2)=&{W_F(z_2)\over z_1-z_2}\,\, .\cr}
\eqn\vseis
$$
$R_F$ and $W_F$ are bosonic operators of dimensions $1$ and $2$
respectively, whose explicit expressions are
$$
\eqalign{
R_F=&\Psi_a\,\overline\Psi_a+\eta_a\lambda_a\cr
W_F=&{1\over \kappa+h}(h\partial\Psi_a\Psi_a-{i\over 2}f^{abc}
\Psi_a\Psi_bJ_c)\,\, .\cr}
\eqn\vsiete
$$
The algebra of $T$, $T_F^{\pm}$, $R_F$ and $W_F$ closes only after the
introduction of a new dimension-${3\over 2}$ operator $V_F$:
$$
V_F={1\over 3\sqrt{2(\kappa+h)}}\,\,f^{abc}\,\,\Psi_a\Psi_b\Psi_c\,\, .
\eqn\vocho
$$
It can be readily verified that $R_F$, $W_F$ and $V_F$ are primary fields.
In fact $V_F$ shows up when acting on $W_F$ with $T_F^-$. One has :
$$
\eqalign{
T_F^{+}(z_1)W_F(z_2)=&0\cr
T_F^{-}(z_1)W_F(z_2)=&{3\over (z_1-z_2)^2}V_F(z_2)+{1\over z_1-z_2}
\partial V_F(z_2)\cr
T_F^{+}(z_1)V_F(z_2)=&{W_F(z_2)\over z_1-z_2}\cr
T_F^{-}(z_1)V_F(z_2)=&0\,\, .\cr}
\eqn\vnueve
$$
The OPE's between $W_F$ and $V_F$ vanish, namely
$$
W_F(z_1)W_F(z_2)=W_F(z_1)V_F(z_2)=V_F(z_1)V_F(z_2)=0\,\, ,
\eqn\treinta
$$
and the product of $R_F$ with itself is non-singular; in fact, $R_F$
introduces a conserved $U(1)$ charge. By inspection one verifies that all
the generators of the extended supersymmetry algebra have a well-defined
$R_F$-charge, which is equal to $+1$, $-1$, $-2$ and $-3$ for $T_F^{+}$,
$T_F^{-}$, $W_F$ and $V_F$ respectively.

We shall call the algebra displayed in eqs. \vseis, \vnueve\ and
\treinta\ the {\it supersymmetric Kazama algebra}; it was first introduced in
\REF\kaza{Y. Kazama \journal\mpl&A6(91)1321.}[\kaza].
In our case this algebra is realized with a vanishing central charge.
Notice that, for $g$  abelian, $W_F$ and $V_F$ vanish
identically and the supersymmetric Kazama algebra is nothing but the
usual $N=2$ superconformal algebra. Moreover the matter and ghost parts
of our model separately realize the supersymmetry algebra with central
charges $3\,\,$dim$\,\,g$ and $-3\,\,$dim$\,\,g$ respectively (actually the
ghost system has an $N=2$ supersymmetry). On the other hand, the form of the
generators in the matter sector displayed in eqs. \vcinco, \vsiete\ and
\vocho\ can be obtained by twisting the realization of the  extended
topological algebras found in refs.[\tca,\gln]
(see also ref.
\REF\getzler{E. Getzler, ``Manin pairs and topological field theory", MIT
preprint(1993) (hep-th/9309057).}[\getzler] where this algebra is termed
$N=1{1\over 2}$ superconformal algebra). The operator $T_F$ that
generates the $N=1$ subalgebra can be obtained as a particular combination
of the three dimension-${3\over 2}$ generators:
$$
T_F={1\over 2}(T_F^{+}+T_F^{-}-{1\over 2}V_F)\,\, .
\eqn\tuno
$$
It can  also be checked that $T_F^{\pm}$ and $R_{F}$ act on the superconformal
currents ${\cal J}_a$ and $\Phi_a$ as follows:
$$
\eqalign{
T_F^{+}(z_1){\cal J}_a(z_2)=&0\cr
T_F^{-}(z_1){\cal J}_a(z_2)=&{\Phi_a(z_2)\over (z_1-z_2)^2}+
{\partial \Phi_a(z_2)\over z_1-z_2}\cr
R_{F}(z_1){\cal J}_a(z_2)=&0\cr
T_F^{+}(z_1)\Phi_a(z_2)=&{{\cal J}_a(z_2)\over z_1-z_2}\cr
T_F^{-}(z_1)\Phi_a(z_2)=&0\cr
R_{F}(z_1)\Phi_a(z_2)=&{-1\over z_{1}-z_{2}}\Phi_a(z_2)\,\, .\cr}
\eqn\tdos
$$
The OPE's of $W_F$ and $V_F$ with the currents vanish. This fact,
together with \tdos, shows the compatibility of the supersymmetric and
current algebras.

Let us now show that this system possesses a BRST symmetry
such that the
currents ${\cal J}_a$ and $\Phi_a$ become BRST-exact. We have at our
disposal two antighost fields $\rho_a$ and $\lambda_a$ with conformal
dimensions $1$ and ${1\over 2}$ respectively, so it's  natural to
require to our BRST transformation $\,\,\delta$ that
$$
\delta \rho_a={\cal J}_a
\,\,\,\,\,\,\,\,\,\,\,\,
\delta \lambda_a=\Phi_a\,\, .
\eqn\ttres
$$
Using eqs. \once, \vdos\ y \vcuatro, it is easy to check that the
transformations in \ttres\
are obtained by acting with the zero mode of the operator
$$
Q=-\gamma_a(J_a+if^{abc}\,\,\overline \Psi_b\Psi_c+
if^{abc}\,\, \lambda_b\eta_c+{i\over 2}f^{abc}\,\, \gamma_b\rho_c)
-i\sqrt {2(\kappa+h)}\,\,\eta_a\Psi_a\,\, ,
\eqn\tcuatro
$$
the $\delta$-variations of the other fields being
$$
\eqalign{
\delta \gamma_a=&{i\over 2}f^{abc}\,\,\gamma_b\gamma_c\cr
\delta \eta_a=&if^{abc}\,\,\gamma_b\eta_c\cr
\delta J_a=&if^{abc}\,\,\gamma_bJ_c+2h\partial \gamma_a\cr
\delta \bar J_a=&if^{abc}\,\,\gamma_b\bar J_c-2(\kappa+h)\partial
\gamma_a\cr
\delta \Psi_a=&if^{abc}\,\,\gamma_b\Psi_c\cr
\delta \overline \Psi_a=&if^{abc}\gamma_b\,\,\overline \Psi_c +
i\sqrt {2(\kappa+h)}\,\,\eta_a\,\, .\cr}
\eqn\tcinco
$$
A standard calculation shows that $Q(z)$ is a nilpotent operator:
$$
Q(z_1)Q(z_2)=0\,\, .
\eqn\tseis
$$
In order to show that $Q$ is a generator of a topological symmetry we
must prove that the energy-momentum tensor $T$ can be written as a BRST
variation. Let us call $G$  the BRST partner of $T$, \ie, the
dimension-$2$ field such that $T=\delta G$. The expression of $G$ is more
conveniently written in terms of the operator
$$
\Lambda_a=\rho_a-{f^{abc}\over\sqrt {2(\kappa+h)}}
\,\,\lambda_b\overline \Psi_c\,\, ,
\eqn\tsiete
$$
which is a dimension-$1$ operator with ghost number
$-1$. Using eqs. \ttres\ and \tcuatro\ it can be readily verified that its
BRST variation is
$$
\delta  \Lambda_a=J_a+if^{abc}\,\,\gamma_b \Lambda_c\,\, .
\eqn\tocho
$$
In terms of $\Lambda_a$, we now define $G$ as
$$
G={1\over 2(\kappa+h)}\bar J_a\Lambda_a +
{i\over 2 \sqrt {2(\kappa+h)}}
(\partial \lambda_a\overline
\Psi_a-\lambda_a\partial\overline\Psi_a)\,\, .
\eqn\tnueve
$$
Indeed, one can check that the OPE of $Q$ and $G$ is
$$
Q(z_1)G(z_2)={T(z_2)\over z_1-z_2}+{R(z_2)\over (z_1-z_2)^2}+{d\over
(z_1-z_2)^3}\,\, ,
\eqn\cuarenta
$$
where
$$
\eqalign{
d=&{\rm dim}\,\,g\cr
R=&\rho_a\gamma_a +{1\over 2}\eta_a\lambda_a+{1\over
2}\overline\Psi_a\Psi_a\,\, .\cr}
\eqn\cuno
$$
Eq. \cuarenta\ is the first in a number of OPE's characterising the topological
symmetry of this model. Other OPE's involving $T$, $Q$, $R$ and $G$ are
$$
\eqalign{
T(z_1)Q(z_2)=&{Q(z_2)\over (z_1-z_2)^2}+{\partial Q(z_2)\over z_1-z_2}\cr
T(z_1)R(z_2)=&-{d\over (z_1-z_2)^3}+{R(z_2)\over (z_1-z_2)^2}+{\partial
R(z_2)\over z_1-z_2}\cr
T(z_1)G(z_2)=&{2G(z_2)\over (z_1-z_2)^2}+{\partial G(z_2)\over z_1-z_2}\cr
R(z_1)R(z_2)=&{d\over (z_1-z_2)^2}\cr
R(z_1)Q(z_2)=&{Q(z_2)\over z_1-z_2}\cr
R(z_1)G(z_2)=&-{G(z_2)\over z_1-z_2}\,\, .\cr}
\eqn\cdos
$$
Notice that, according to eq.\cdos, $Q$ and $G$ are primary fields with
dimensions $1$ and $2$ respectively, whereas $R$ is an anomalous $U(1)$
current, $d$ being the corresponding anomaly. We shall call $d$ the
{\it dimension of the topological algebra}; in our case $d$ equals precisely
the dimension of the Lie algebra $g$. Curiously, the same value of $d$ is
obtained in the non-supersymmetric topological current algebras [\tca].

An important feature of the topological algebra just described is the fact that
$G$ is {\it not} a nilpotent operator. Actually the OPE of $G$ with itself
gives rise to a new dimension-$3$ operator $W$ :
$$
G(z_1)G(z_2)={W(z_2)\over z_1-z_2}\,\, ,
\eqn\ctres
$$
where $W$ is a commuting field given by
$$
W={1\over [2(\kappa+h)]^2}\,\,
(if^{abc}J_a\Lambda_b\Lambda_c-2h\partial\Lambda_a\Lambda_a)\,\, .
\eqn\ccuatro
$$
This $W$ operator  is BRST-exact: its BRST ancestor is
$$
V={i\over 3[2(\kappa+h)]^2}\,\,\,f^{abc}\Lambda_a\Lambda_b\Lambda_c\,\, ,
\eqn\ccinco
$$
which is an anticommuting dimension-$3$ operator. Acting with
$Q$ on $V$ one gets a simple pole with $W$ as residue:
$$
Q(z_1)V(z_2)={W(z_2)\over z_1-z_2}\,\, .
\eqn\cseis
$$
After the introduction of these two new fields, the topological algebra
generated by $Q$, $R$, $G$, $T$, $V$ and $W$ closes. Apart from those already
displayed in eqs. \cuarenta, \cdos , \ctres\ and \cseis, the
non-vanishing OPE's are
$$
\eqalign{
R(z_1)W(z_2)=&-{2\over z_1-z_2}\,\,W(z_2)\cr
T(z_1)W(z_2)=&{3\over (z_1-z_2)^2}\,\,W(z_2)+{\partial W(z_2)\over
z_1-z_2}\cr
G(z_1)W(z_2)=&{3\over (z_1-z_2)^2}\,\,V(z_2)+{\partial
V(z_2)\over z_1-z_2}\cr R(z_1)V(z_2)=&-{3\over z_1-z_2}\,\,V(z_2)\cr
T(z_1)V(z_2)=&{3\over (z_1-z_2)^2}\,\,V(z_2)+{\partial V(z_2)\over
z_1-z_2}\,\, .\cr}
\eqn\csiete
$$
Therefore the topological symmetry of our model is generated by three BRST
doublets of operators (($R,Q$), ($G,T$) and ($V,W$)) of dimensions $1$, $2$
and $3$. It is important to point out that the topological algebra we
have obtained is {\it not} the twisted $N=2$ algebra; in fact what we have
 is an extended topological algebra that can be obtained by
twisting a supersymmetric Kazama algebra of the type described above. We
have obtained another realization of this extended topological symmetry in
refs. [\tca,\gln]  in our study of the topological conformal field
theories  possessing a bosonic,
non-abelian current algebra (see also [\getzler]). The $G/G$
coset theories
\REF\witGG{E. Witten \journal\cmp&144(92)189.}
\REF\yank{M. Spiegelglas and S. Yankielowicz
\journal\np&393(93)301.}
\REF\aharo{O. Aharony et al.\journal\np&B399(93)527
 \journal\pl&B289(92)309 \journal\pl&B305(93)35.}
\REF\hu{H.L. Hu and M. Yu \journal\pl&B289(92)302
\journal\np&B391(93)389.}[\witGG,\yank,\aharo,\hu]
are particular cases of these models. It is interesting to observe
that the same algebra appears when one requires  the topological theory
to have a superconformal current symmetry.

Let us now study the compatibility of the topological symmetry and the
superconformal current algebra. By inspecting the form of the topological
$U(1)$ current (\ie, $R$ in eq. \cuno) one immediately finds out the
charges of the fields:
$\Psi_a$,$\overline \Psi_a$, $\rho_a$, $\gamma_a$, $\lambda_a$ and $\eta_a$
have $R$-charges equal to ${1\over 2}$,$-{1\over
2}$,$-1$,$1$,$-{1\over 2}$ and  ${1\over 2}$, respectively, whereas the bosonic
currents $J_a$ and $\bar J_a$ are neutral with respect to $R$. This means
that the superconformal currents $(\Phi_a,{\cal J}_a)$ and their BRST
ancestors $(\lambda_a,\rho_a)$ have well-defined $R$-charge ($({1\over
2},0)$ for the $(\Phi_a,{\cal J}_a)$ currents and $(-{1\over
2},-1)$ for their topological partners $(\lambda_a,\rho_a)$). Moreover the
OPE's of $G$ with  $(\Phi_a,{\cal J}_a)$ and
  $(\lambda_a,\rho_a)$  are
$$
\eqalign{
G(z_1)\Phi_a(z_2)=&{1\over 2}{\lambda_a(z_2)\over (z_1-z_2)^2}+
{\partial \lambda_a(z_2)\over z_1-z_2}\cr
G(z_1){\cal J}_a(z_2)=&{\rho_a(z_2)\over (z_1-z_2)^2}+
{\partial \rho_a(z_2)\over z_1-z_2}\cr
G(z_1)\lambda_a(z_2)=&G(z_1)\rho_a(z_2)=0\,\, ,\cr}
\eqn\cocho
$$
which confirms our previous conclusion that the currents have the right
transformation properties under the topological symmetry:
the singular expansions of eq. \cocho\ are the ones expected for the
products of the topological partner of the energy-momentum tensor and the
superconformal currents. On the other hand, after some calculation one can
conclude that  no singularity appears when the dimension-$3$ operators
$W$ and $V$ are multiplied by the  superconformal currents.
Altogether this implies that the currents are primary with respect
to the whole set of generators of the topological algebra and therefore
the topological and supercurrent symmetries are indeed compatible.

We finally turn our attention to the relationship between the topological
 and supersymmetry
structures of our model. First of all we notice that the supersymmetry
generators $T_F^{+}$, $T_F^{-}$, $R_F$, $V_F$ and $W_F$ all possess a
well-defined topological $U(1)$ quantum number: they transform under
$R$ as fields with charges $-{1\over 2}$, ${1\over 2}$, $0$, ${3\over 2}$
and $1$ respectively. However, the $N=1$
supersymmetry generator $T_F$ does not transform as an eigenstate of the
$R$-current. Indeed, a glance at eq. \tuno\ reveals that $T_F$
splits into three contributions, each of them with a well-defined
$R$-charge and which are precisely the  generators of the extended
supersymmetry algebra. This fact means that the topological symmetry can
be considered as responsible for the enlargement of the supersymmetry that
 is produced at the topological point of the matter + ghost system.
Moreover, all  the generators of the extended supersymmetry are BRST trivial,
\ie\ there exist new fields ${\cal T}_F^{\pm}$, ${\cal R}_F$, ${\cal
W}_F$ and ${\cal V}_F$ such that
$$
\eqalign{
T_F^{\pm}=&\,\,\,\delta {\cal T}_F^{\pm}
\,\,\,\,\,\,\,\,\,\,\,\,\,\,\,\,\,\,\,\,\,
R_F=\delta {\cal R}_F\cr
W_F=&\,\,\,\delta {\cal W}_F
\,\,\,\,\,\,\,\,\,\,\,\,\,\,\,\,\,\,\,\,\,
V_F=\delta {\cal V}_F\,\, .\cr}
\eqn\cnueve
$$
The explicit expressions for the new fields appearing in eq. \cnueve\ are
$$
\eqalign{
{\cal T}_F^{+}=&-{i\over 2 \sqrt{2(\kappa+h)}}\,\,
\overline \Psi_a(\rho_a+\Lambda_a)\cr
{\cal T}_F^{-}=&{1\over 2(\kappa+h)}\,\,\lambda_a\bar J_a\cr
{\cal R}_F=&-{i\over \sqrt{2(\kappa+h)}}\,\,\overline\Psi_a\lambda_a\cr
{\cal W}_F=&-{1\over [2(\kappa+h)]^{3\over 2}}\,\,
\Psi_a(f^{abc}\,\,J_b\lambda_c+2ih\partial\lambda_a)\cr
{\cal V}_F=&-{i\over 6(\kappa+h)}\,\,
f^{abc}\,\,\Psi_a\Psi_b\lambda_c\,\, .\cr}
\eqn\cincuenta
$$
The BRST-exactness of the supersymmetry generators implies that all
states created by acting on the vacuum with a finite number of them
 can be gauged away. In other words, the extended supersymmetry
of our model is topological, and   the cohomology of $Q$ encodes the
global information about the supersymmetric Hilbert space.

In order to achieve a better understanding of the relationship between the
supersymmetric and topological symmetries of our model, let us study how
the BRST current $Q$ acts on $T^{\pm}_F$ and $R_F$. A straightforward
calculation gives the result
$$
\eqalign{
Q(z_1)T_F^{+}(z_2)=&0\cr
Q(z_1)T_F^{-}(z_2)=&-{I(z_2)\over (z_1-z_2)^2}\cr
Q(z_1)R_F(z_2)=&0\,\, ,\cr}
\eqn\newuno
$$
where $I$ is a new dimension-${1\over 2}$ bosonic current whose explicit
expression is
$$
I=i\sqrt{2(\kappa +h)}\gamma_a\Psi_a+{i\over
2}f^{abc}\gamma_a\gamma_b\lambda_c\,\, .
\eqn\newdos
$$
Moreover, when $Q$ acts on ${\cal T}_F^{\pm}$ and ${\cal R}_F$, their
topological partners are obtained as residue of the simple pole
singularity,  while some extra contributions appear in the double pole.
One has:
$$
\eqalign{
Q(z_1){\cal T}_F^{+}(z_2)=&{T_F^{+}(z_2)\over z_1-z_2}\cr
Q(z_1){\cal T}_F^{-}(z_2)=&{{\cal I}(z_2)\over (z_1-z_2)^2}+
{T_F^{+}(z_2)\over z_1-z_2}\cr
Q(z_1){\cal R}_F(z_2)=&-{d\over (z_1-z_2)^2}+
{R_F(z_2)\over z_1-z_2}\,\, ,\cr}
\eqn\newtres
$$
where $d$ is the dimension of the topological algebra (\ie,
$d={\rm dim}\,\,g$ as in eqs. \cuarenta-\cdos), and ${\cal I}$ is a
dimension-${1\over 2}$ fermionic field given by
$$
{\cal I}=-\gamma_a\lambda_a\,\, .
\eqn\newcuatro
$$
It can be readily checked that ${\cal I}$ is the BRST ancestor of $I$, \ie,
$$
Q(z_1){\cal I}(z_2)={I(z_2)\over z_1-z_2}
.\eqn\newcinco
$$
We thus see that, in  trying to relate $T_F^{\pm}$ and $R_F$ to their BRST
partners, we are forced to introduce a new BRST doublet of fields $({\cal I},
I)$. Actually, a whole plethora of new operators appear when the
generators of the
topological algebra, on the one hand, and those of the
 supersymmetric one, on the other, are multiplied together. The reason for
this proliferation of operators can be traced back to the non-nilpotency
of one of the generators of the extended supersymmetric algebra (\ie\ of
$T_F^{-}$). Indeed one would expect that all operators appearing in the
symmetry algebra could be arranged into  supersymmetry multiplets whose
components are generated by acting with $T_F^{+}$ and $T_F^{-}$. As
$(T_F^{-})^2\not= 0$, many new components of the supersymmetry multiplet
are generated in this process. However when $g$ is taken to be an
abelian algebra (\ie, when $f^{abc}=h=0$), this problem disappears since
our system possesses an $N=2$ supersymmetry. One can then   easily check
in this case  that the full algebra closes with
the sole addition of $I$ and its BRST partner ${\cal I}$ to the
generators of the topological and supersymmetry algebras.

Let us consider now the question of the uniqueness of our construction.
We could try to modify the matter sector by considering more
matter species (\ie, by taking $M>2$ in eqs. \dsiete -\dnueve). It is
immediate to see that, in this case, the vanishing of the central charge $c$
does not follow from the zero level condition $x=0$. Of course we
could adjust the levels $k_i$ in  such a way that $c=0$, but we  still must
demand the BRST exactness of ${\cal J}_a$ and $\Phi_a$ (see eq. \ttres)
and of $T$. This latter condition means that we outght to  be able to find an
operator $G$ such that $T=\delta G$. It can be easily concluded that, when
$M>2$, it is not possible to find an expression for $G$ local in the
currents. The proof of this statement is the same as in the bosonic case
(see ref. [\tca]) and will not be reproduced here. This result implies
that our construction only works for $M=2$. At this point it is
interesting to recall [\tca] that the bosonic topological current algebras
can be defined for $M=1,2$, contrary to what happens in the present
supersymmetric case in which $M=1$ is not allowed.

The realization of the extended topological symmetry we have found admits
deformations, \ie, redefinitions of its generators
such that the transformed operators satisfy the same
algebra. Suppose that $\alpha_a$ are a set of c-number constants, and let
us redefine $T$, $G$ and $R$ as follows:
$$
\eqalign{
T&\rightarrow T+\sum_a\alpha_a\partial {\cal J}_a\cr
G&\rightarrow G+\sum_a\alpha_a\partial \rho_a\cr
R&\rightarrow R+\sum_a\alpha_a {\cal J}_a\,\, ,\cr}
\eqn\ciuno
$$
while $Q$, $V$ and $W$ are not transformed. It's not difficult to see that the
redefined fields still close the extended topological algebra for the
same value of the parameter $d$ (\ie\ for $d=\,\,$dim$\,\,g$). Of course,
this transformation does not preserve the supercurrent symmetry, but
it does preserve the supersymmetric Kazama algebra if we redefine
$T_F^{\pm}$ and $R_F$ as
$$
\eqalign{
T_F^{+}&\rightarrow T_F^{+}\cr
T_F^{-}&\rightarrow T_F^{-}+2\sum_a\alpha_a\partial\Phi_a\cr
R_F&\rightarrow R_F+2\sum_a\alpha_a{\cal J}_a\,\, .\cr}
\eqn\cidos
$$

Let us now recapitulate our main results. We have been able to find a
topological conformal model possessing a superconformal current algebra
compatible with the topological symmetry, in which the generators of
the current
algebra are exact in the BRST cohomology defined by the topological
symmetry. The topological algebra closes only after the introduction of
two dimension-$3$ operators ($W$ and $V$ in  eqs. \ccuatro\ and \ccinco).
The compatibility of the supersymmetry with the topological symmetry of
the model requires the extension of the former. The generators of the
extended supersymmetric algebra include three dimension-$3$ fermionic
operators ($T_F^{\pm}$, $V_F$), a dimension-$1$ $U(1)$ current ($R_F$)
and a dimension-$2$ bosonic field ($W_F$). All these generators are BRST
exact, a fact which exhibits the topological nature of the supersymmetric
algebra.

Our results raise several questions that deserve further study. First of
all, we would like to point out that the model we have constructed is the
supersymmetric analogue of the $G/G$ coset model. The states of the $G/G$
topological field theory are intimately connected with the conformal
blocks of the Wess-Zumino-Witten theory for the group $G$ [\yank.\aharo],
 so it's natural to conjecture that the states of our model are in
correspondence with the conformal blocks of the supersymmetric WZW model.
Furthermore it has been shown [\yank,\aharo,\hu] that, after a deformation
such as the one in eq. \ciuno, the physical states of the theory are
equivalent to those of the minimal models coupled to gravity (\ie, to
those of the non-critical string theory). Therefore, in our case, one would
expect to get a relationship of the deformed theory with the non-critical
superstrings. On the other hand, the SL(N)/SL(N) models have been obtained in
ref. [\gln] from the WZW model based on the GL(N,N)  supergroup. From this
fact one is led to suspect that the model constructed above could be described
as a suitable supersymmetrization of the  GL(N,N) WZW theory. We expect to
analyse these topics in the near future.

\ack
The authors would like to thank H.L. Hu, J.M.F. Labastida, P.M.
Llatas, J.~Mas, and J.~S\'anchez de Santos for discussions. One of us (J.M.I.)
is grateful to the Department of Physics of Queen Mary and
Westfield College, where the last part of this work was carried out, for
hospitality, and to  Conselleria de Educacion da Xunta de Galicia for
financial support. This work was supported in part by DGICYT under grant
PB90-0772, and by CICYT under grants  AEN90-0035 and AEN93-0729.

\endpage

\refout
\end